\documentclass[a4paper,11pt]{article}

\usepackage{jinstpub} 
\usepackage{siunitx}
\usepackage{subfigure}
\usepackage{lineno}
\pdfoutput=1

\title{Discharge and stability studies for the new readout chambers of the upgraded ALICE TPC}

\author[a,b,1]{A. Deisting,\note{Corresponding author.}}
\author[a]{C. Garabatos,}

\affiliation[a]{GSI Helmholtzzentrum f\"ur Schwerionenforschung GmbH\\ Darmstadt, Germany}
\affiliation[b]{Physikalisches Institut, Ruprecht-Karls-Universit\"at Heidelberg\\ Heidelberg, Germany}

\emailAdd{alexander.deisting@cern.ch}

\abstract{The ALICE (A Large Ion Collider Experiment) Time Projection Chamber (TPC) at CERN LHC is presently equipped with Multi Wire Proportional Chambers (MWPCs). A gating grid prevents ions produced during the gas amplification from moving into the drift volume. The maximum drift time of the electrons together with the closure time of the gating grid allows a maximum readout rate of about \SI{3}{\kilo\hertz}. After the Long Shutdown 2 (from 2021 onwards), the LHC will provide lead-lead collisions at an expected interaction rate of \SI{50}{\kilo\hertz}. To take data at this rate the TPC will be upgraded with new readout chambers, allowing for continuous read-out and preserving the energy and momentum resolution of the current MWPCs.\\
Chambers with a stack of four Gas Electron Multipliers (GEMs) fulfil all the performance requirements, if the voltages applied to the GEMs are tuned properly. In order to ensure that these chambers are stable while being operated at the LHC, studies of the discharge behaviour were performed. We report on studies done with small prototypes equipped with one or two GEMs. Discharges were voluntarily induced by a combination of high-voltages across the GEM(s) and highly ionising particles. During these studies, the phenomenon of "secondary discharges" has been observed. These occur only after an initial discharge when the electric field above or below the GEM is high enough. The time between the initial and the secondary discharge ranges from several $\SI{10}{\micro\second}$ to less than \SI{1}{\micro\second}, decreasing with increasing field. Using decoupling resistors in the high-voltage supply path of the bottom side of the GEM shifts the occurrence of these discharges to higher electric fields.}


\arxivnumber{1705.02150} 

\collaboration[c]{on behalf of the ALICE collaboration}

\proceeding{INSTR17: Instrumentation for Colliding Beam Physics\\
  27 February - 3 March, 2017\\
  Novosibirsk, Russia}

\begin{document}
\maketitle
\flushbottom

\section{The ALICE TPC and the ALICE TPC Upgrade}
\label{sec:alicetpc}

ALICE (A Large Ion Collider Experiment) is one of the four experiments at the CERN Large Hadron Collider (LHC) and is dedicated to the study of heavy-ion collisions. ALICE uses a Time Projection Chamber (TPC) for tracking and charged-particle identification in the central barrel.\\
The ALICE TPC\cite{alme2010alice} is a \SI{5}{\metre} long cylindrical detector, with the readout chambers located at the endplates, facing the central (drift) cathode in the middle. Hence the drift volume is split into two volumes with a drift length of \SI{2.5}{\meter} each. The chambers themselves are Multi-Wire Proportional Chambers (MWPCs) with a gating grid used to block ions produced in the readout chambers from moving into the drift volume. The closure time of the gating grid ($\sim\!\SI{250}{\micro\second}$), together with the maximum time an electron needs to drift to the readout chambers ($\sim\!\SI{90}{\micro\second}$), imposes an upper limit on the readout rate of about \SI{3}{\kilo\hertz}.\\ \indent
After Long Shutdown 2, the LHC will provide lead-lead collisions at \SI{50}{\kilo\hertz}. To exploit the full interaction rate, the TPC will be upgraded. The ALICE TPC Upgrade project developed new readout chambers using stacks of four gas electron multipliers (GEMs)\cite{sauli1997gem}, which allow for continuous readout and preserve the performance as achieved with the current TPC\cite{aliceTpcUpgradeTDR2014,aliceTpcUpgradeTDR2015Addendum}. Each readout chamber will be equipped with a GEM stack containing two Large-Pitch (LP) GEM foils (hole pitch of \SI{280}{\micro\meter}) at position 2 and 3 in the stack and two Standard pitch (S) foils (hole pitch of \SI{140}{\micro\meter}) at the position 1 and 4. The GEMs themselves will be segmented on the side facing the drift volume and unsegmented on the bottom side. Each of this segments has an area of approximately $10\times\SI{10}{\centi\meter\squared}$ and will be biased through a \SI{5}{\mega\ohm} loading resistor.\\ \indent
Different studies of the discharge probability conducted for the ALICE TPC Upgrade are described in \cite{aliceTpcUpgradeTDR2015Addendum}. During these tests, the High-Voltage (HV) settings foreseen for the upgraded TPC were applied to a quadruple GEM stack under the baseline gas mixture $\textrm{Ne}$-$\textrm{CO}_{2}$-$\textrm{N}_2$ (90-10-5). A discharge probability of \SI[separate-uncertainty = true]{6(4)e-12}{discharges} per incoming hadron was found. Only a small number of discharges is expected in the upgraded TPC from the measured discharge probability, considering that \SI{7e11}{particles} are expected to hit each GEM stack in the TPC during one month of lead-lead collisions.

\section{Studies of the discharge mechanism}
\label{sec:dcstudies}
\subsection{Experimental set-up}
\label{sec:dcstudies:subsec:setup}
\begin{figure}[tb]
\centering
\includegraphics[width=.75\textwidth,trim=0 20 0 60,clip]{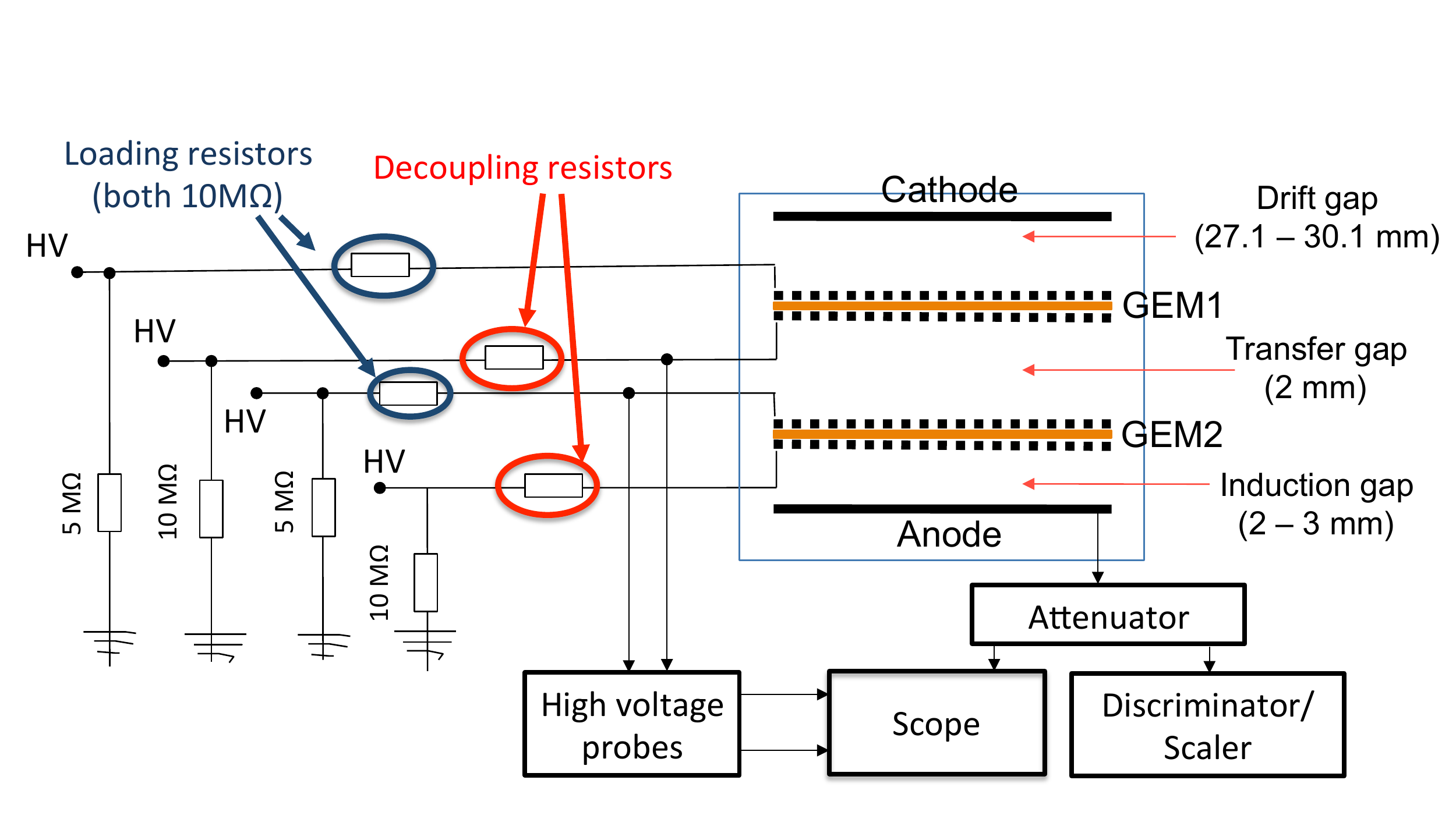}\caption{\label{fig:doubleGEMsetup} A sketch of the set-up used for the discharge studies: In the detector one or two $10\times\SI{10}{\centi\meter\squared}$ GEMs are mounted. The counting gas is enriched with Radon and discharges are induced by a combination of high-voltages across the GEMs and ionisations created by the $\alpha$ decays of $\textrm{Rn}$. Two high-voltage probes are used at different electrodes to investigate the potentials during a discharge.}
\end{figure}

The aim of this study is to analyse the behaviour of GEM foils during discharges. Figure~\ref{fig:doubleGEMsetup} illustrates the set-up used for the discharge measurements and shows its main features. Measurements are performed with LP and S GEMs of $10\times\SI{10}{\centi\meter\squared}$ size, either operated as single-GEM or as a stack of two GEMs. An $\textrm{Ar}$-$\textrm{CO}_{2}$ (90-10) mixture at atmospheric pressure is flushed through a container with Thorium mantels. The $^{222}\textrm{Rn}$ isotopes from the $^{230}\textrm{Th}$-decays are carried by the gas flow into the detector and, as they undergo an $\alpha$ decay cause a discharge in (one of) the GEM foil(s) with a given probability.\\ \indent
All GEM electrodes are powered with a CAEN 470 Power Supply (PS) with four independent channels. On the top side of each GEM a \SI{10}{\mega\ohm} loading resistor is placed. The bottom sides of the foils were either powered directly or with decoupling resistors. In order to sink excess currents in case of a discharge, a resistor to ground was added to each HV channel. This resistor is not present in the HV scheme of the ALICE TPC. To enhance the discharge rate, the voltage across the GEM(s) ($\Delta U_{\textrm{GEM}}$) is moderately increased.
The discharge in a GEM induces a signal on the anode plane, which is then attenuated by a \SI{10}{\kilo\ohm} resistor in series with a \SI{3}{dB} T-type attenuator. Afterwards the signal is fed into a discriminator and, if higher than a set threshold ($\sim\SI{150}{\milli\volt}$), counted by a scaler. A signal from the discharge needs $\mathcal{O}(\SI{50}{\micro\second})$ to decay (see figure~\ref{fig:anaodesecondary}). To prevent double counting of the same discharge, a gate of several \SI{100}{\micro\second} was used. No dead-time was imposed by this method, as the GEM takes $\mathcal{O}(\SI{100}{\milli\second})$ to recharge due to the RC-constant of the supply circuit.\\ \indent
The signals are recorded by an oscilloscope and stored for later analysis. Two high-voltage probes are used in addition to examine the potentials on the GEM sides during a discharge. Each probe consisted of a $\SI{345}{\mega\ohm}$ resistor in parallel to $22\times\SI{1.5}{\pico\farad}$ capacitors in series. During a measurement the probe is connected to the electrode of interest and to the oscilloscope. The probes draw therefore a small current since they are connected to ground via the input resistance of the scope. In some configurations, mainly when measuring the potentials on the top side of the GEMs, this current leads to a voltage drop. In such cases, the applied voltages are adjusted to compensate for the expected voltage drop. Since the response of both probes depends on the amplitude of the original signal, a proper calibration with direct and alternating current signals was performed. It was also checked that the use of the probes at a GEM did not alter its discharge probability compared with the same settings without probes. 

\subsection{Discharge probability}
\label{sec:dcstudies:subsec:dcprob}
\begin{figure}[tb]
\centering
\subfigure[]{
\includegraphics[width=.45\textwidth,trim=0 0 0 55,clip]{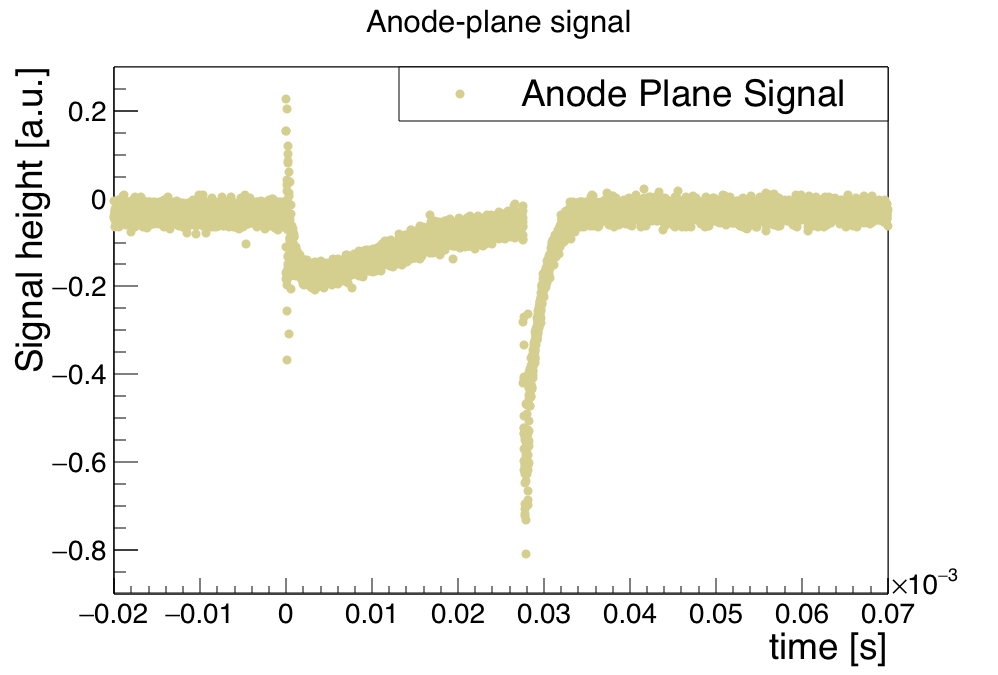}
\label{fig:anaodesecondary}}
\subfigure[]{
\includegraphics[width=.45\textwidth,trim=0 0 0 55,clip]{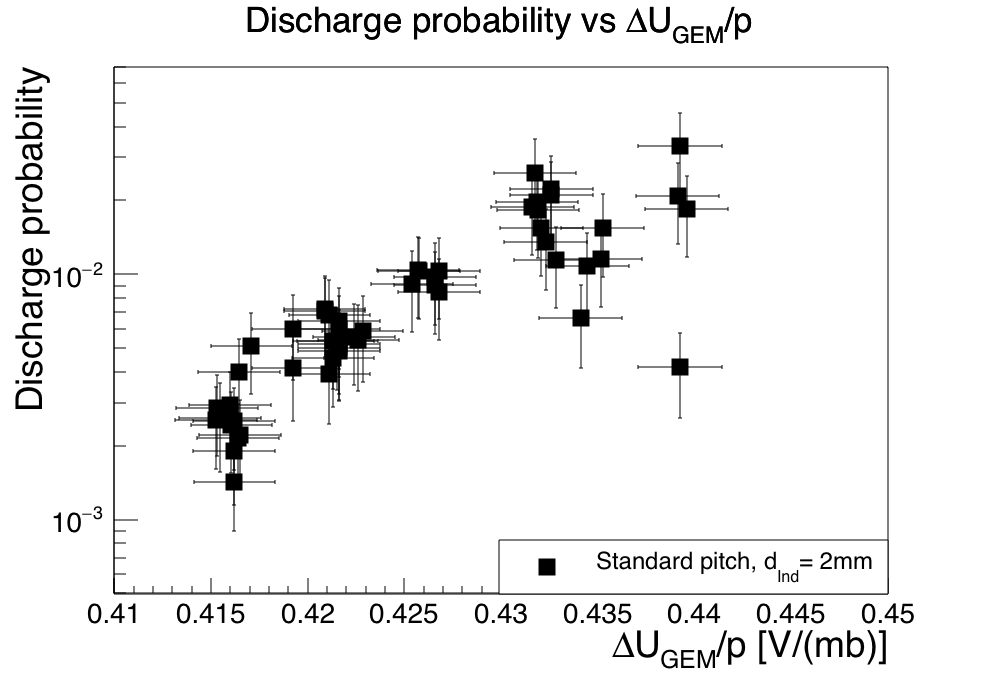}
\label{fig:dcprob}}
\caption{\label{fig:generaldc}\protect\subref{fig:anaodesecondary}: Signal obtained during a discharge in the GEM close to the anode plane: The discharge appears at $t=\SI{0}{\second}$ followed by a secondary discharge at $t\sim\SI{28}{\micro\second}$. \protect\subref{fig:dcprob}: Probability for an $\alpha$ decay of $^{222}\textrm{Rn}$ to result in a discharge in a single-GEM set-up as a function of the reduced voltage of one GEM, in $\textrm{Ar}$-$\textrm{CO}_{2}$ (90-10) with a water content between \SI{125}{ppm} and \SI{175}{ppm}. The length of the drift and the induction gap were $d_{\textrm{Drift}}=\SI{27.6}{\milli\meter}$ and $d_{\textrm{Ind}}=\SI{2}{\milli\metre}$, respectively. No decoupling resistor was used for the measurements displayed in both plots.}
\end{figure}
The decay rate of the $\alpha$-source was measured by signal counting without attenuator but preamplifier at the anode plane. From this and the measured discharge rate, the probability for a decay to trigger a discharge was calculated as $P_{1} = \frac{\textrm{discharge}\ \textrm{rate}}{\textrm{decay}\ \textrm{rate}}$. Figure~\ref{fig:dcprob} shows $P_{1}$ for a measurement with a single S GEM. The discharge probability increases with increasing $\Delta U_{\textrm{GEM}}$. This is driven by the increasing charge density in the GEM holes as recently modelled in \cite{gasik2017}, which grows with the gas gain. To disentangle the ambient pressure dependence, all given fields (voltages) are divided by the pressure.

\subsection{Secondary discharges}
\label{sec:dcstudies:subsec:secondaries}
While performing discharge studies with single- and multi-GEM structures, a second type of discharge is occasionally observed after an initial discharge. In a setting without decoupling resistor, these "secondary discharges" are characterised by an amplitude several times higher than that of the original discharge (see figure~\ref{fig:anaodesecondary}).
\begin{figure}[tb]
\centering
\subfigure[]{
\includegraphics[width=.45\textwidth,trim=0 0 0 55,clip]{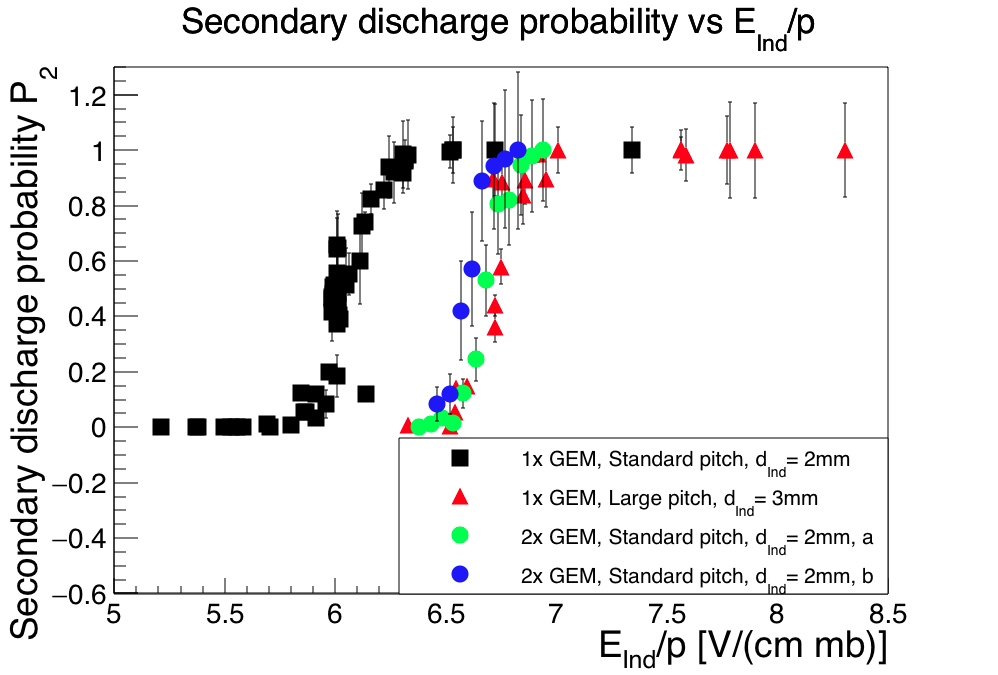}
\label{fig:p2vseind}}
\subfigure[]{
\includegraphics[width=.45\textwidth,trim=0 0 0 55,clip]{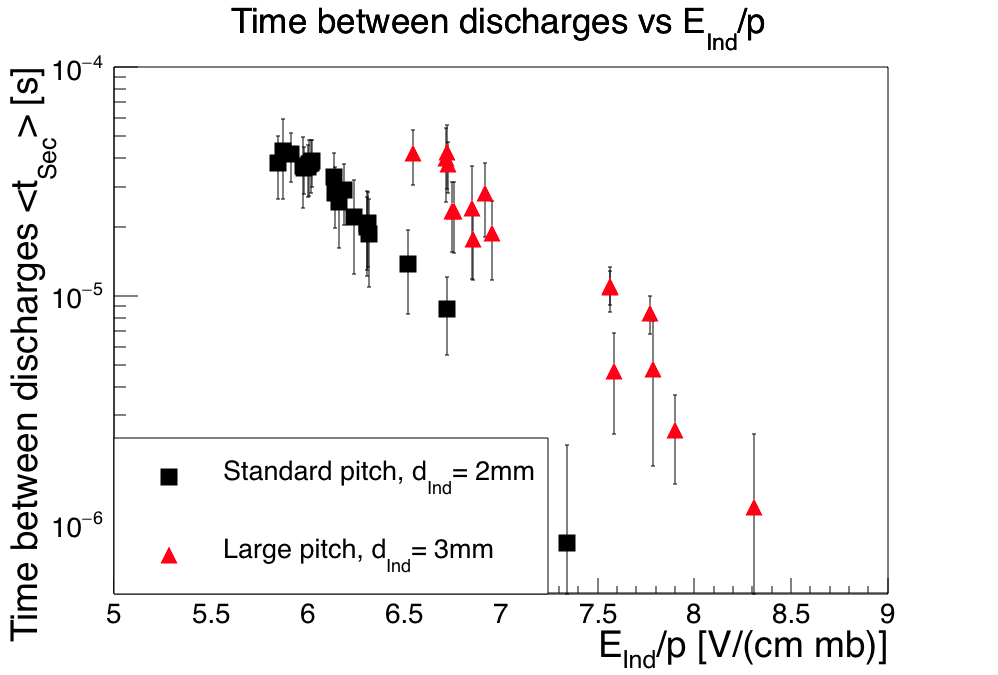}
\label{fig:t12vseind}}
\caption{\label{fig:sndvseind} Secondary discharges as a function of the reduced induction field. Measurements are shown for a large-pitch and a standard GEM in a single-GEM configuration, and for different standard GEMs in a double-GEM set-up. In all cases the GEM close to the anode plane discharged.  Different GEMs at the position close to the anode plane are represented by different symbols. No decoupling resistor was used. \protect\subref{fig:p2vseind}: Probability $P_2$ for a secondary discharge to occur after the initial discharge. The two data sets indicated with full circles (a, b) differ by the GEM foil used as GEM1. \protect\subref{fig:t12vseind}: Average time $\langle t_{\textrm{Sec}} \rangle$ between initial and secondary discharge for the single-GEM measurements displayed in figure \protect\subref{fig:p2vseind}.}
\end{figure}
Similar phenomena have been reported in \cite{bachmann2002discharge} and \cite{peskov2009research} ("fully propagated" and "delayed" discharges) it was, however, only realised in recent studies for the ALICE TPC Upgrade\cite{pgasik2016rd51} that the occurrence of secondary discharges depends on the field below the GEM.\\ \indent
In order to study these secondary discharges, the counting logic described previously was extended by another discriminator, gate and scaler. The signal from the anode plane is split and fed into two discriminators with different thresholds. A discharge signal higher than the respective threshold would open the corresponding gate and would be counted as either initial or as secondary discharge. No precondition such as the presence of a previous discharge is imposed on the counting of the secondary discharges. An analysis of the counts and recorded waveforms showed that secondary discharges are only observed after an initial discharge. The risk of counting secondary discharges as the initial ones is avoided, since the gate length of $\mathcal{O}(\SI{100}{\micro\second})$ was longer than the longest time difference between the first and secondary discharge.\\ \indent
Figure~\ref{fig:sndvseind} shows four measurement series conducted with both a single-GEM and a double-GEM configuration. There, the induction field ($E_{\textrm{Ind}}$) is increased stepwise while the potential differences between all the other electrodes are kept constant. The secondary discharge probability in figure~\ref{fig:p2vseind} is calculated as $P_{2} = \frac{\#\;\textrm{secondary}\ \textrm{discharges}}{\#\;\textrm{initial}\ \textrm{discharges}}$, while the error bars are the statistical errors on the counts. Figure~\ref{fig:t12vseind} shows the average time between the initial and the secondary discharge $\langle t_{\textrm{Sec}} \rangle$ for different $E_{\textrm{Ind}}$. Times $t_{\textrm{Sec}}$ were extracted for every secondary discharge from the anode-plane signals stored during the discharge counting and binned into a histogram. This is done for each voltage setting. The different $\langle t_{\textrm{Sec}} \rangle$ correspond to the mean values of the $t_{\textrm{Sec}}$-distributions in these histograms. The corresponding RMS is chosen as error.\\ \indent
No dependence of the secondary discharges on the $\Delta U_{\textrm{GEM}}$ is identified. With the double-GEM set-up secondary discharges in the transfer gap are observed as a function of the transfer field $E_{\textrm{T}}$. The onset curve has the same shape as curves measured in the induction gap, but the rise of $P_{2}$ from 0 to 1 takes place for smaller values of $E_{\textrm{T}}$. This is illustrated by the data points denoted by circles in figure~\ref{fig:p2vseind} and by the $R_{\textrm{Dec}}=0$ points in figure \ref{fig:sndvstransdec}. Both measurements were recorded with the same settings. In case of the latter plot, $E_{\textrm{T}}$ is stepwise increased and $E_{\textrm{Ind}}=\SI{1}{\kilo\volt\per\centi\meter}$ is kept.

\subsection{Usage of decoupling resistors}
\begin{figure}[tb]
\centering
\subfigure[]{
\includegraphics[width=.45\textwidth,trim=0 0 0 55,clip]{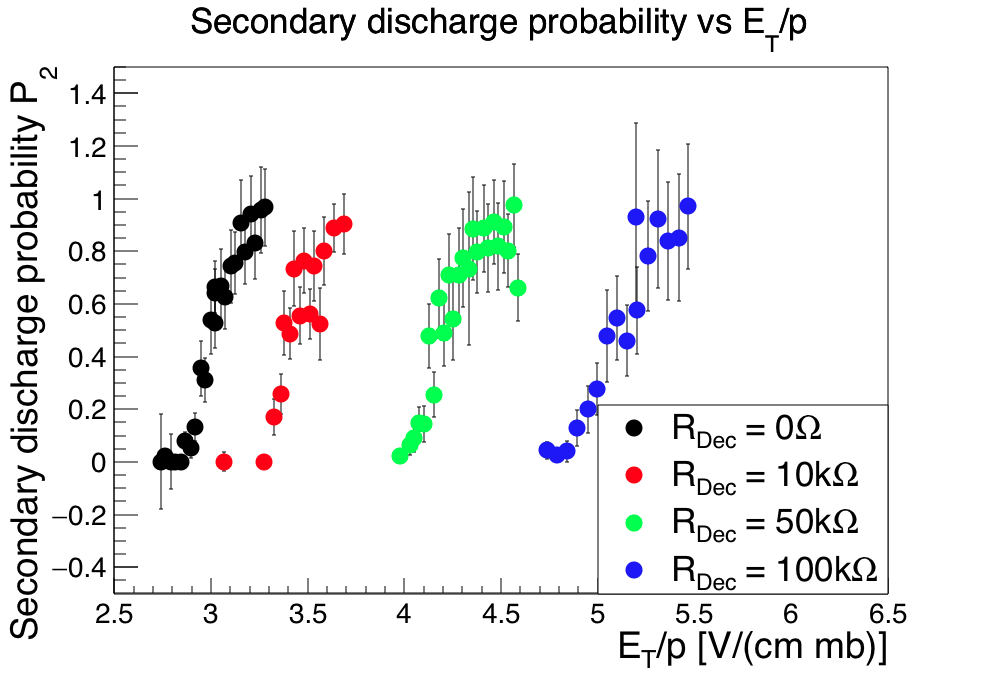}
\label{fig:sndvstransdec}}
\subfigure[]{
\includegraphics[width=.45\textwidth,trim=0 0 0 25,clip]{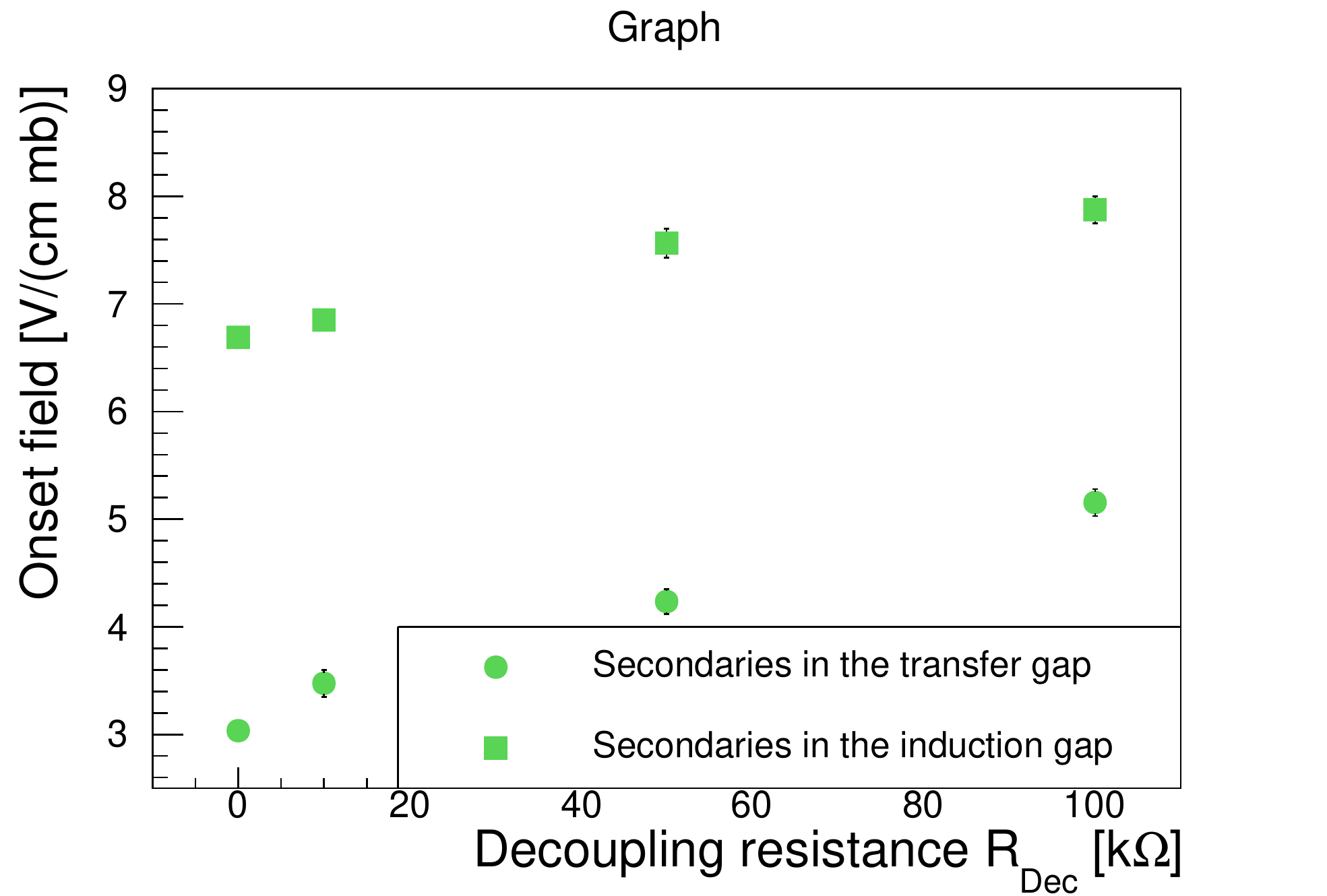}
\label{fig:onsetvsres}}
\caption{\label{fig:secondwithres} With a double-GEM set-up (two standard GEMs) the probability to obtain a secondary discharge in the transfer (induction) gap was studied as a function of the transfer field $E_{\textrm{T}}$ (induction field $E_{\textrm{Ind}}$) and decoupling resistance\cite{baDatz2017}. \protect\subref{fig:sndvstransdec}: $P_2$ (analogue to figure~\ref{fig:p2vseind}) as a function of the transfer field is plotted for different decoupling resistors. \protect\subref{fig:onsetvsres}: Comparison of the electric field at which $P_2$ reaches 0.5 ("onset field") for the measurements shown in \protect\subref{fig:sndvstransdec} and for a similar measurement with secondary discharges in the induction gap.}
\end{figure}
Different measurements with decoupling resistor, as well as the corresponding measurement without, are displayed in figure~\ref{fig:secondwithres}. The higher the resistance, the higher the electric field at which secondary discharges are observed. In figure~\ref{fig:onsetvsres} the "onset-field" ($E_{\textrm{onset}}$) of secondary discharges is defined as the field at which $P_2 = 0.5$. For secondary discharges in the induction gap\cite{pgasik2016res,baDatz2017} as well as in the transfer gap\cite{baDatz2017}, the corresponding onset-fields followed a linear trend as a function of the decoupling resistance.

\subsection{Evolution of potential during the different discharges}
\label{sec:potentials}
\begin{figure}[hb]
\centering
\subfigure[]{
\includegraphics[width=.45\textwidth,trim=0 0 0 50,clip]{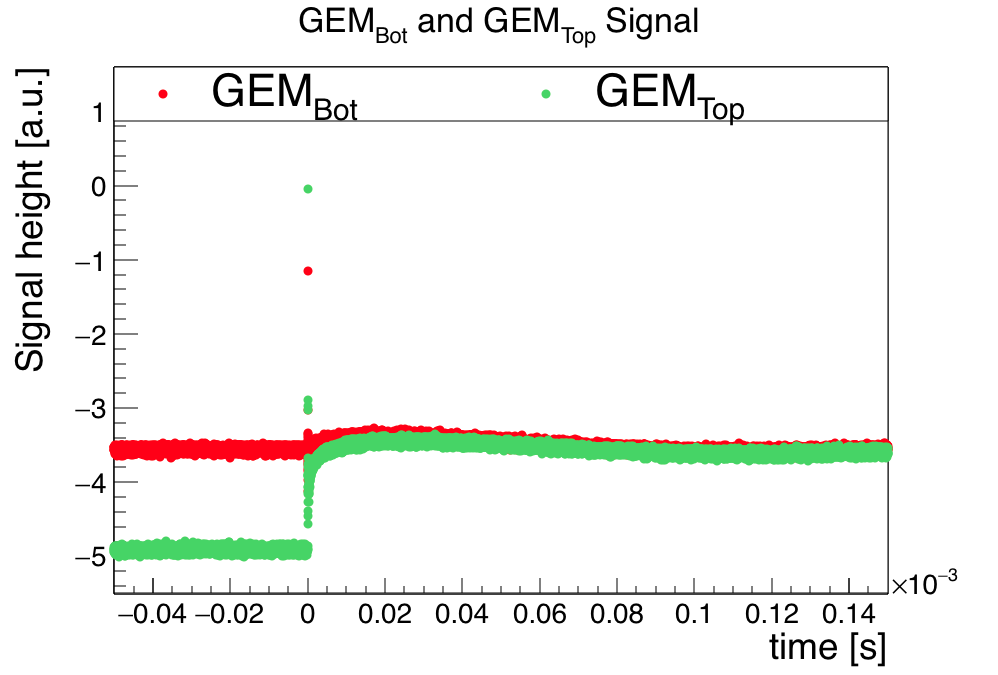}
\label{fig:probeSignalwithout}}
\subfigure[]{
\includegraphics[width=.45\textwidth,trim=0 0 0 50,clip]{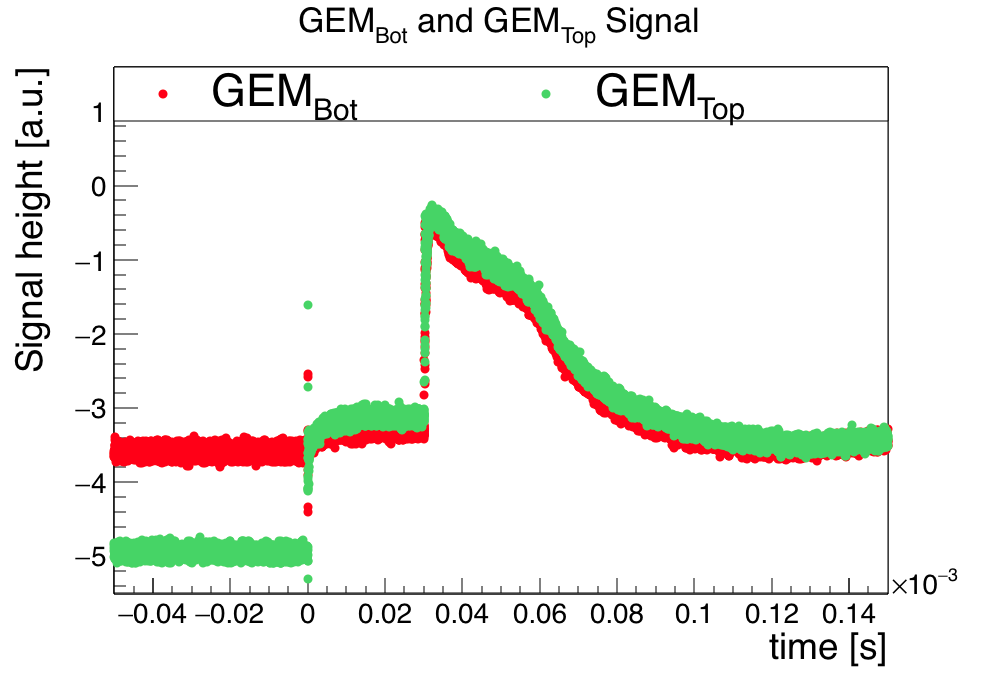}
\label{fig:probeSignalwith}}
\caption{\label{fig:probeSignal} Waveforms recorded with the HV probes at the top and bottom side of a GEM. As indicated in figure~\ref{fig:doubleGEMsetup} probes are connected directly at the GEM and therefore, as seen from the HV supply, after the loading resistor. These waveforms were obtained in measurements with the single-GEM set-up and no decoupling resistor. \protect\subref{fig:probeSignalwithout}: A discharge at $t=0$. \protect\subref{fig:probeSignalwith}: A discharge at $t=0$ followed by a secondary discharge at $t\sim\SI{30}{\micro\second}$.}
\end{figure}

To monitor the $\Delta U_{\textrm{GEM}}$ of the discharging GEM, high-voltage probes are connected to the top and bottom sides of the foil. At the oscilloscope the potentials appear downscaled by the voltage divider of the probe resistance and the input resistance of the oscilloscope. Since only negative potentials are applied, more negative amplitude values in the waveforms correspond to higher potentials. Waveforms in figure~\ref{fig:probeSignal} are recorded, while probing the potentials of the GEM close to the anode plane. Only a discharge (\ref{fig:probeSignalwithout}) and a discharge followed by a secondary discharge (\ref{fig:probeSignalwith}) are displayed. Directly after the discharge, potentials show fast bipolar oscillations with a high amplitude for about $\mathcal{O}{(\SI{1}{\micro\second}})$. Because of the oscilloscope settings used to record the data in figure~\ref{fig:probeSignal}, these oscillations are only visible as few points that do not follow the general trend of the respective signals. In the anode plane signal in figure~\ref{fig:anaodesecondary} oscillations are visible as well. They are here superimposed with an unipolar signal, which is most likely due to movement of charge released during the discharge. At the same time, the potential on the top side of the GEM foil drops to the bottom potential as the GEM discharges, while the potential on the bottom side remains roughly unchanged. This is due to the loading resistor on the top side of the GEM foil. If there is a decoupling resistor, the bottom potential rises and both potentials settle at an intermediate value between the original potentials present before the discharge. From the analysis of all the recorded potentials the voltage across the GEM after a discharge is found to be at most $\mathcal{O}(\SI{10}{\volt})$ and compatible with zero, considering the error bars on the probe measurement. 
The secondary discharges were observed to occur as $\Delta U_{\textrm{GEM}}\sim0$. Both potentials drop immediately towards the potential at the anode plane, which is at ground potential as shown in figure~\ref{fig:probeSignalwith}. Here as well, oscillations of the potentials occur. The qualitative picture is the same for secondary discharges in the transfer gap of the double-GEM set-up. The potential across GEM2 (the GEM close to the anode plane) drops as well during the discharge. During the secondary discharge, however, the two potentials on the top and bottom side of GEM2 increase. At the same time, both potentials of GEM1 drop. The potential values at the top side of GEM2 and the bottom side of GEM1 suggest that both potentials reach the same absolute value for a short time.\\ \indent
While the recovery of a secondary discharge happens relatively fast ($\mathcal{O}(\SI{100}{\micro\second})$), the recharging of the GEM foil takes several hundreds of \si{\milli\second} as expected by the RC-constant of the GEM and the loading resistor.

\section{Discussion}
\label{sec:dis}
We observe that the GEM potentials approach the potential of the anode plane during a secondary discharge (secondary discharge in the induction gap) and that the potentials of the two GEM sides facing the transfer gap approach the same value (secondary discharge in the transfer gap). These suggest that the secondary discharge shortens the gap for a very limited time. The time evolution of the measured potentials between initial and secondary discharges are the same, regardless of whether the latter occurs. Hence no process that could explain the discharge of the gap has been identified. Furthermore, no increase of $\Delta U_{\textrm{Gap}}$ was measured for the induction gap before the secondary discharge takes place. The potential across the transfer gap increases by the potential drop on GEM2. Any other potential increase is not observed. The initial discharge of the GEM appears to be a necessary condition of the secondary discharge.\\ \indent
In all the recorded data series (figure~\ref{fig:p2vseind} and \ref{fig:sndvstransdec}) $P_2$ exhibits a steep onset from zero to one as  observed in ref~\cite{pgasik2016rd51}. Comparing the onset fields of measurements with different GEMs (measurements without decoupling resistor), one obtains \SI[separate-uncertainty=true]{6.05(10)}{\volt\per\centi\meter\per\milli\bar} for the standard GEM and \SI[separate-uncertainty=true]{6.73(10)}{\volt\per\centi\meter\per\milli\bar} for the large-pitch GEM in the single-GEM configuration. A different standard GEM at the GEM2 position in the double-GEM configuration yields $E_{\textrm{Ind}\ \textrm{Onset}} = \SI[separate-uncertainty=true]{6.63(10)}{\volt\per\centi\meter\per\milli\bar}$. While the latter onset is close to the one observed with the LP GEM, there is a significant deviation from the onset value observed for the standard GEM, used in the single-GEM set-up. This indicates that the process responsible for the creation of the secondary discharge slightly differs for different GEMs. Comparing the two single-GEM data series in figure~\ref{fig:p2vseind} with $\langle t_{\textrm{Sec}} \rangle$ in figure~\ref{fig:t12vseind} allows us to conclude that a different onset field results in a shift in the average time between initial and secondary discharges. From the reduced ion mobilities reported in \cite{Deisting2017215} the velocity of $\textrm{CO}_{2}$ ions in $\textrm{Ar}$-$\textrm{CO}_{2}$ (90-10) can be calculated as \SI{15}{\micro\second} at an electric field of \SI{6.5}{\kilo\volt\per\centi\meter}. Measured durations between primary and secondary discharge however decreases from several \SI{10}{\micro\second} to less than \SI{1}{\micro\second} for an electric field in the range of \SI{6.2}{\kilo\volt\per\centi\meter} to \SI{8.2}{\kilo\volt\per\centi\meter}. This and the fact that ions are mostly produced close to the GEM suggest that ions crossing the gap are not responsible for the secondary discharge. The delay between initial and secondary discharge rules out a photon induced effect creating the secondary discharge. Explanations relying on charges created by photons of the initial discharge are again not compatible with the observed $\langle t_{\textrm{Sec}} \rangle$ and the expected drift times between the electrodes.\\ \indent
The results of the studies with the different decoupling resistors show that $E_{\textrm{Onset}}$ can be moved to significantly higher values for higher decoupling resistance. This may be related to the different behaviour of the GEM potentials during a discharge when GEMs are equipped with a loading resistor. In both cases, running with or without decoupling resistor, the potentials on both sides of the GEM are affected in case of a secondary discharge in the induction gap. In case of a secondary discharge in the gap between two GEMs, all the potentials of these two GEMs will change. These potential changes can result into an increase of the fields between GEMs in a GEM stack, leading to a series of multiple (secondary) discharges as observed while inducing secondary discharges in the induction gap and operating the double-GEM set-up with a high transfer field. The GEM potentials change less during a secondary discharge when the GEM is equipped with a decoupling resistor

\section{Summary}
\label{sec:sum}
Dedicated studies of the discharge behaviour of GEM foils were performed in $\textrm{Ar}$-$\textrm{CO}_{2}$ (90-10) with a single- and with a double-GEM set-up. Discharges were induced in GEMs by combining an $\alpha$-source and high-voltage across the respective GEM. The anode plane signal as well as two potentials were recorded. It was found for the potential across a GEM to be compatible with zero after a discharge. Only the potential on the top side of the GEM drops, if operated at this side with a loading resistor of \SI{10}{\mega\ohm}.\\ \indent
Secondary discharges were also observed. This type of discharges occurs in all observations only after an initial discharge when the field below or above the GEM is high enough. They appeared to happen from several \SI{10}{\micro\second} down to less than \SI{1}{\micro\second} after the discharge. This time is found to decrease if the field used to trigger them is increased. It was observed that the gap adjacent to the GEM is shortened during this second type of discharge as the secondary discharge is induced by a high electric field across this gap. The potentials on both sides of a GEM are affected during a secondary discharge. The onset curve of the secondary discharges is observed to take place over only a very small electric field range. Different onset fields were found for different GEMs. Values between \SI[separate-uncertainty=true]{6.05(10)}{\volt\per\centi\meter\per\milli\bar} and \SI[separate-uncertainty=true]{6.73(10)}{\volt\per\centi\meter\per\milli\bar} were measured for secondary discharges in the induction gap. For the transfer gap an onset field of \SI[separate-uncertainty=true]{3.03(10)}{\volt\per\centi\meter\per\milli\bar} was found. During studies using the baseline gas mixture for the ALICE TPC Upgrade ($\textrm{Ne}$-$\textrm{CO}_{2}$-$\textrm{N}_2$ (90-10-5)) the onset of secondary discharges in the transfer gap was observed at lower values as compared to measurements with $\textrm{Ar}$-$\textrm{CO}_{2}$ (90-10) as counting gas. A similar observation was made in \cite{pgasik2016rd51} for secondary discharges in the induction gap. \\ \indent
The introduction of a decoupling resistor at the high-voltage path of the bottom side of the GEM foil resulted in an increase of the onset field. Since the transfer fields of the ALICE TPC baseline high voltage-settings will be as high as \SI{4}{\kilo\volt\per\centi\meter}, the introduction of such resistors seems promising to make the occurrence of secondary discharges unlikely.\\ \indent
However, the mechanism resulting in the occurrence of secondary discharges is not yet understood.


\end{document}